# The International Space Station (ISS) Contest as STEM Educational Project

### By Enzo Bonacci[*]

*In the years 2015–2018, the Italian Ministry of Education, University and Research and the Italian Ministry of Defense proposed the joint initiative "Space for Your Future. The ISS: Innovatio, Scientia, Sapientia" in partnership with the Italian Space Agency. It was a competition addressed to secondary school students and aimed at developing innovative experiments to be conducted on the International Space Station, whose acronym (ISS) is the same of the Latin words "Innovatio, Scientia, Sapientia". Regardless of the odds of winning, "Space for Your Future" became a successful STEM educational project implemented in numerous schools. We illustrate how that Astronomy contest fostered a valid constructivist learning, a fruitful participatory science, and vast scientific research. We discuss, in particular, the activities of two teams of pupils from the Scientific High School "Giovanni Battista Grassi" in Latina (seat of the Planetarium "Livio Gratton") who participated within the thematic area No. 3 "Test the Sciences in Space". They all worked on chemical tests, suitable for the ISS microgravity, under the tutoring of Francesco Giuliano (Province Manager of the IYA 2009 and the IYC 2011 in Latina). The key reference is a talk given in the 104th annual congress of the Italian Physical Society at the University of Calabria (September 17–21, 2018) together with an invited lecture held in the 13th European Researchers' Night by Frascati Scienza (September 28, 2018).*

**Introduction**

Launched jointly by the Italian Ministry of Education, University and Research (MIUR for brevity) and by the Italian Ministry of Defense (MDI for short), the contest "Space for Your Future. The ISS: Innovatio, Scientia, Sapientia" had a slow start in 2015 (Figure 1), despite the collaboration with the Italian Space Agency (ASI). Only after a persuasion campaign led by the Italian cosmonaut Lieutenant Colonel Walter Villadei, the MIUR-MDI-ASI's initiative was understood in its ambitious purpose of connecting the School to Space research and welcomed as a big opportunity to improve the teenagers' attitude towards STEM (Science, Technology, Engineering and Mathematics). The Italian cosmonaut elucidated how the Latin words "Innovatio, Scientia, Sapientia", translatable as "Innovation, Science, Wisdom", had been accurately chosen for both their meaning and acronym (ISS) identical to the famous International Space Station. He explained that each secondary school could compete with maximum three teams, picked from the 4th and 5th year's pupils (ages 17–19) and preferably steered by mentors from accredited institutions.

---

[*]Teacher, formerly at Scientific High School "G.B. Grassi" of Latina, Italy.

**Figure 1.** *The 2015 Official Poster of the ISS Competition for Italian Schools*

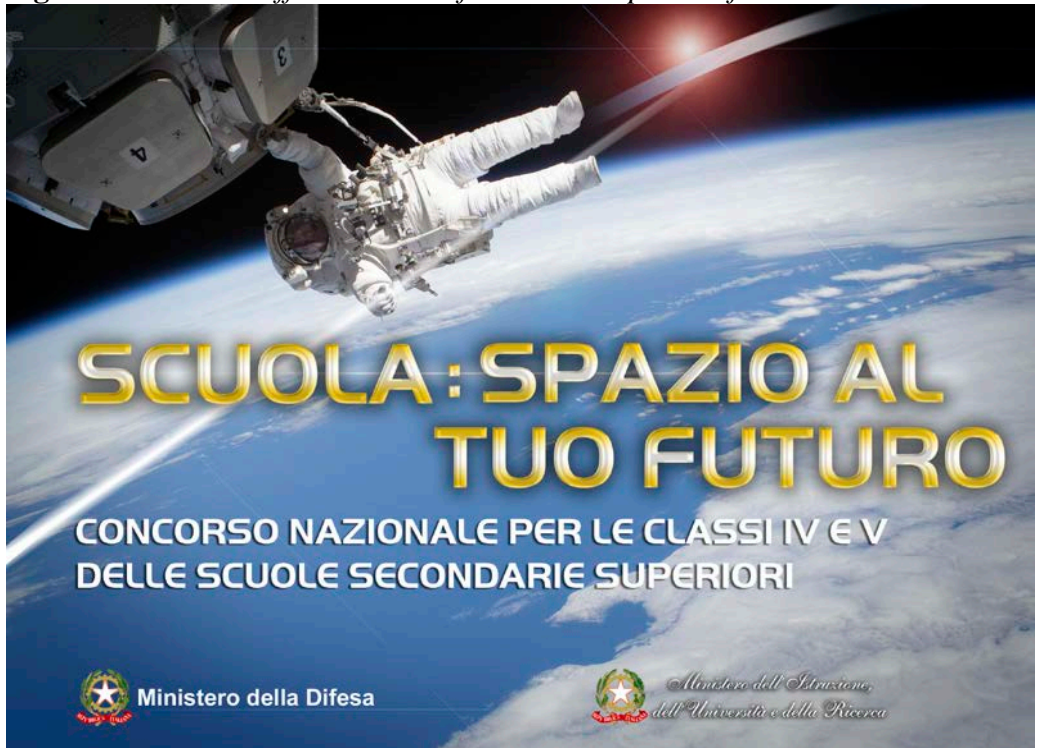

*Source*: https://bit.ly/3qMJYzJ.

Walter Villadei clarified that some of the selected works would possibly be tested in microgravity conditions right on board the ISS (orbiting 400 kilometres above the Earth). In his tour across Italy (Fig. 2), he expounded that "Space for Your Future" was conveniently divided into the following seven thematic areas:

1. "From daily life to Space": proposals to improve the lives of astronauts aboard the International Space Station.
2. "Train like an astronaut": proposals for the development of physical exercises, sensors, and tools to improve the efficiency of physical exercise on board, also through new protocols and new strategies usable for the training of astronauts in a simulated situation on earth.
3. "Test the Sciences in Space": proposals for activities and/or experimental protocols of natural sciences, physics, chemistry, and biology aimed at learning about the space environment and also aimed at highlighting the differences between the space and terrestrial environment.
4. "Observe the Earth to guard it": proposals for the development of computer applications (App) related to Earth observation, the development of thematic catalogs, new observational techniques, or the use of innovative tools to monitor our habitat, protect it and preserve it.

5. "Stay connected with an astronaut": proposals for the development of computer applications (App) aimed at interacting directly with astronauts and bringing the life of the astronaut on board the Space Station closer to the daily life of students on the ground.
6. "Robots, satellites and astronauts conquering the Universe": proposals for the development of microsatellite prototypes, automatic systems interacting with humans, prototypes and/or advanced robotics experiments on board the ISS.
7. "Cultivate in space to cultivate better on Earth": project proposals for the development of new cultivation techniques in the space environment that can also provide information on how to use available resources in hostile environments.

**Figure 2.** *Walter Villadei Presents the ISS Contest in Latina (May 27, 2016)*

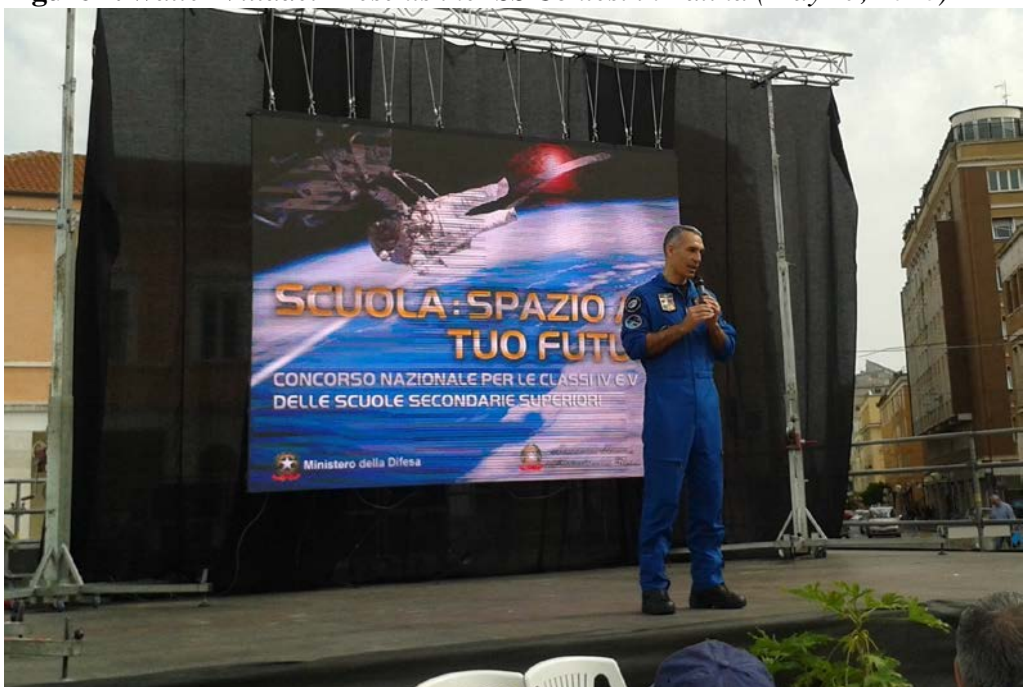

*Source*: https://bit.ly/3CfzL2b.

Fifty-four Italian schools joined the *ISS* competition through STEM educational projects based on constructivist methods, principally Problem and Project Based Learning (PBL & PrBL), which enhanced the Inquiry Based Science Education (IBSE), the Education for Sustainable Development (ESD), the Engineering Design Process (EDP), and the Information & Communication Technologies (ICT). Here we offer an overview of the competition, from the first call for applications (December 15, 2015) to the awarding ceremony (March 14, 2018), with a brief mention to the forerunner SUCCESS student contest by ESA[1]. Later we focus on the contribution from the Scientific High School "G.B. Grassi" of Latina (Lazio region), whose pupils formulated two chemical experiments in mini containers. Most of the material about that inspiring STEM activity comes from a topical talk (Bonacci, 2018a) and a public lecture (Bonacci, 2018b) meant to be rigorous (Heigl et al., 2019), accessible and engaging, as auspicated by the European Citizen Science Association[2]. The legal framework and an up-to-date literature are retrieved from current institutional and sectorial websites.

---
[1]https://bit.ly/3LXGQcz.

**The Trailblazing Student Contest SUCCESS by ESA**

As outlined in the Fifty-fourth session of the United Nations Committee on Peaceful Uses of Outer Space (COPUOS): "Since 1998, when the first modules were launched, the ISS has been assembled by a partnership of five space agencies (CSA, ESA, JAXA, NASA, and Roscosmos) representing 15 countries. The unique features of the ISS are: robust, continuous, and sustainable microgravity platform; continuous human presence in space; access to the ultra-high vacuum of space; unique altitude for observation and testing; and payload-to-orbit-and-return capability."[3] As soon as the International Space Station was ready, the European Space Agency (ESA) ran the *Space station Utilisation Contest Calls for European Student initiativeS* (SUCCESS): "a competition for European university students from all disciplines to propose an experiment that could fly on board the ISS."[4] The SUCCESS contest acquired a fair notoriety in 2005, when a selected experiment was really conducted on board the International Space Station (Fig. 3). ESA's SUCCESS paved the way for similar competitions such as the Italian *Space for Your Future* (targeted on secondary schools) we examine in the next paragraph.

**Space for Your Future - The ISS: Innovatio, Scientia, Sapientia**

*The ISS First Call for Applications in 2015 (Deadline March 31, 2016)*

On December 15, 2015, the Italian Ministry of Education, University and Research launched "the competition *Space for Your Future. The ISS: Innovatio, Scientia, Sapientia* addressed to pupils of fourth and fifth years of high schools, technical and professional institutes. Through the competition, aimed at the dissemination and promotion of scientific activities and technologies in the space sector, students will be involved in the conception of innovative experimentation proposals to be taken aboard the International Space Station. Students will be asked to develop experimentation proposals (actual artifacts and/or experimentation protocols), to be carried out on board the International Space Station, based on the technical and scientific skills acquired during the schooling period and subsequently elaborated with the support of the teachers. and the sponsoring organizations of the initiative" (translated from the MIUR Note No. 13482, Italian website[5]).

---

[2]https://bit.ly/3juKZJ5.
[3]https://bit.ly/3oQRtVN.
[4]https://bit.ly/3zdcAD3.
[5]https://bit.ly/3AyfEv1.

**Figure 3.** *Hardware of Bone Proteomics Flown on the ISS in April 2005*

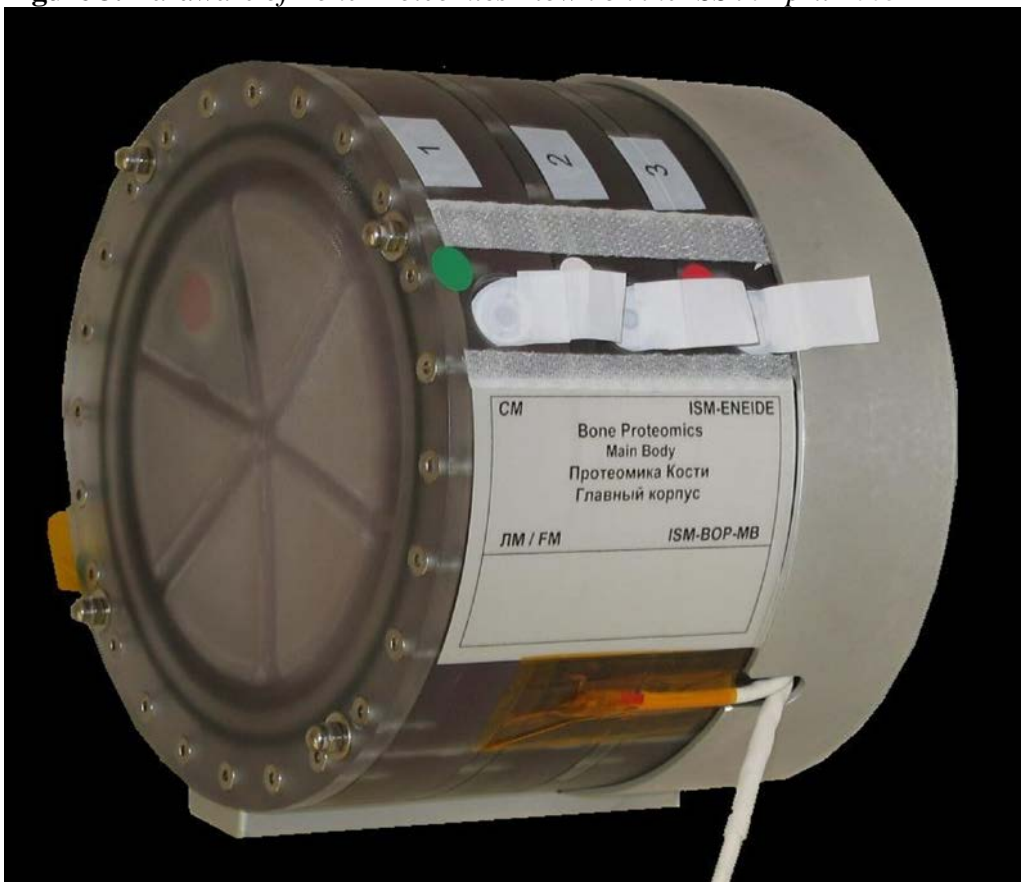

*Source*: https://bit.ly/40I9ZNb.

*The ISS Second Call for Applications in 2016 (Deadline March 31, 2017)*

On September 16, 2016, the MIUR revised the flyer (Figure 4) and some aspects of the regulation, included the contest's deadline (extended until March 31, 2017) and the description of the definitive project:

"Scope the final project represents the conclusive work (final report) prepared by the teams joining the "School: space for your future" initiative and which will be presented for subsequent evaluation by the Commission set up for the Competition Announcement. The Final Projects must contain proposals for experiments to be carried out in space and/or on board an orbiting platform (primarily ISS) within the thematic areas described in the Competition Rules.

Requirements the *Final Project* must meet some requirements:

- expose contents consistent with the premises proposed in the preliminary project, in terms of thematic areas and proposed objectives;
- report plainly the working methods adopted;
- describe the various logical, experimental and investigation phases of the works;

- report in detail the national or international bibliographic sources referred in the works;
- report a minimum structure including: "Index of topics", "Materials and Methods", "Activities carried out", "Conclusions".

Specific chapters functional to evidence the activities carried out, with a logical-consequential structure, may then be added.

**Figure 4.** *The 2016 Official Poster of the ISS Competition for Italian Schools*

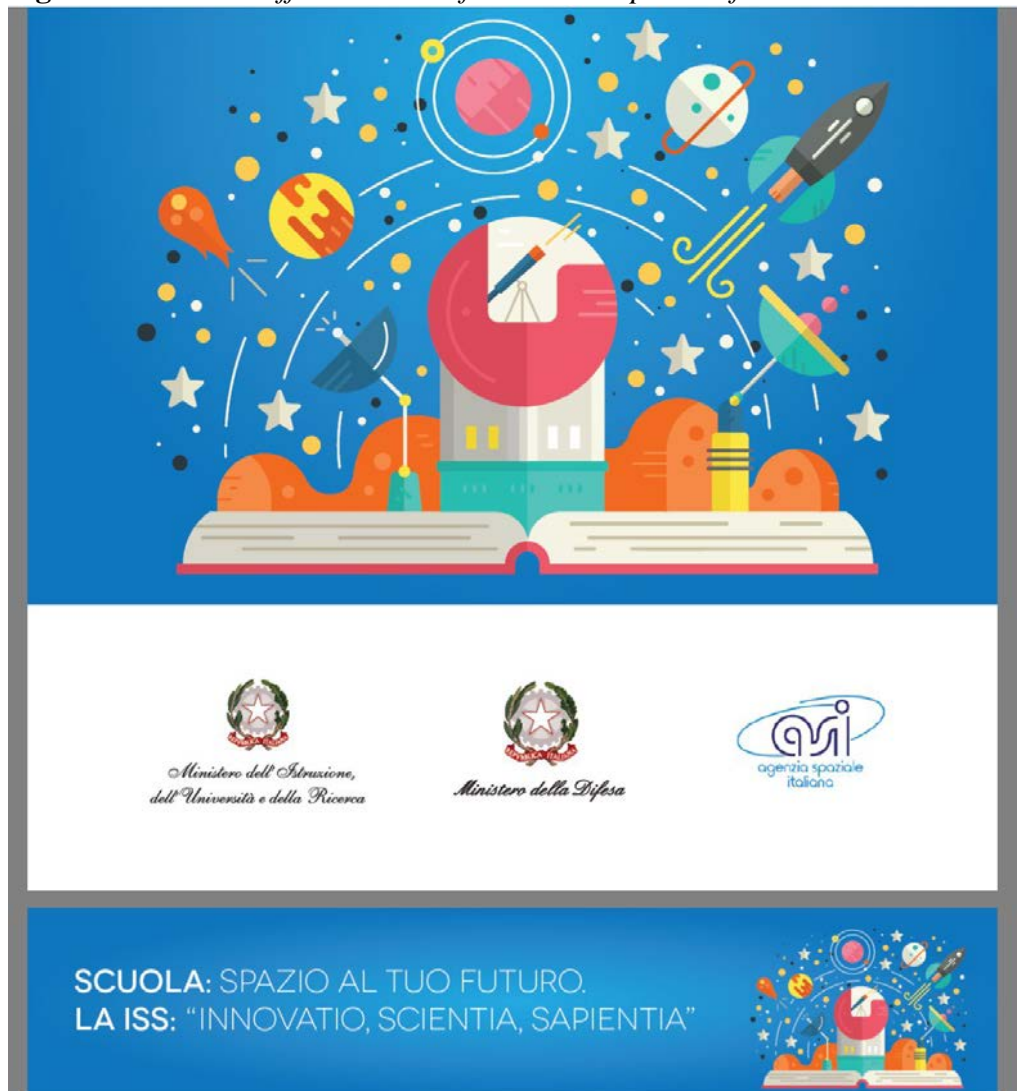

*Source*: https://bit.ly/468Cl6E.

<u>Working methods</u> different kind of *Definitive Projects* can be elaborated: compilation, theoretical, numerical-analytical, experimental. In the case of numerical-experimental projects it will be necessary to report any algorithms and the conceptual structure of the paper. In the case of experimental projects, an accurate and documented description of what has been done will be required. The use of aids such as graphics, images, diagrams, videos will be allowed in order to give evidence of the envisaged results. Any demonstrators or artefacts (technological or scientific) exemplifying the activity carried out and/or the results achieved will be admitted, provided they are complete with adequate explanations. Simulation tools based on commercial software or developed by students using the coding languages they know are also admitted." (translated from the MIUR Note No. 10475, Attachment No.3, Italian website[6]).

*The ISS Third Call for Applications in 2017 (Deadline May 31, 2017)*

On February 23, 2017, the Italian Ministry of Education, University and Research recorded 110 *preliminary* projects[7] presented by 54 schools from 12 regions (Fig. 5) and 33 provinces (Fig. 6). Hence, on March 30, the MIUR announced that "following the numerous requests received from schools and in order to allow more time in the development and processing of experimentation projects, also considering their complexity, the deadline for the submission of the related projects, already set for March 31, 2017, is extended to May 31, 2017" (translated from the MIUR Note No. 3480, Italian website[8]).

Let us notice how the *ISS* challenge was faced positively at all latitudes of the Italian peninsula and in Sicily (Fig. 6).

Since the Italian secondary schools respond to local administrations (regional or provincial, depending on their curricula), we have arranged the geographic distribution of the presented projects in Table 1. The Province of Latina (agro-industrial hub in southern Lazio) joined the *ISS* competition with two secondary schools: the Institute of Higher Education "San Benedetto" and the Scientific High School "G.B. Grassi". The latter[9] hosts the Planetarium "Livio Gratton", a mighty driver of STEM projects (Bonacci, 2011b, 2016a, 2016b) and participatory science (Bonacci, 2017c, 2020).

---

[6]https://bit.ly/3VzYynH.
[7]https://bit.ly/3ChUsus.
[8]https://bit.ly/3AfAmi6.
[9]https://bit.ly/443wGNk.

**Figure 5.** *The 2017 Italian Regional Map of the ISS Registered Projects*

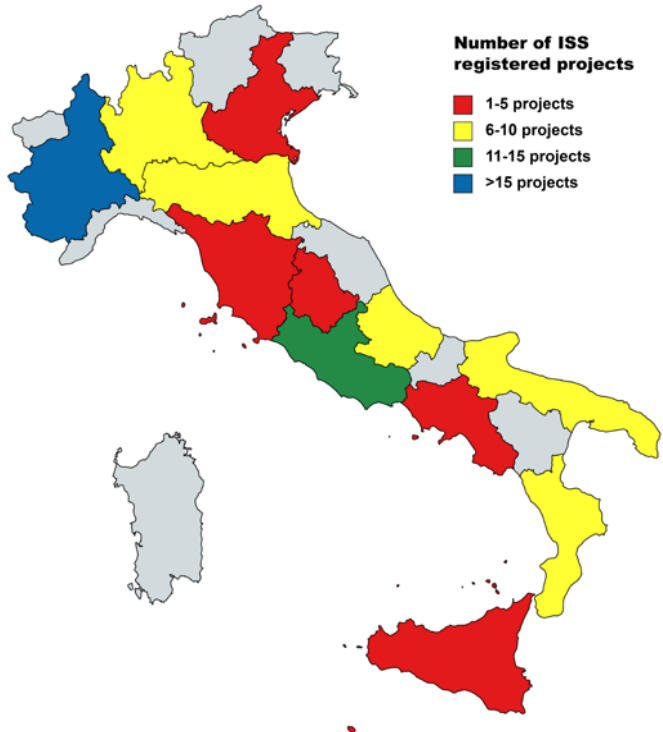

*Source*: www.mapchart.net/europe-detailed.html.

**Figure 6.** *The 2017 Italian Provincial Map of the ISS Registered Projects*

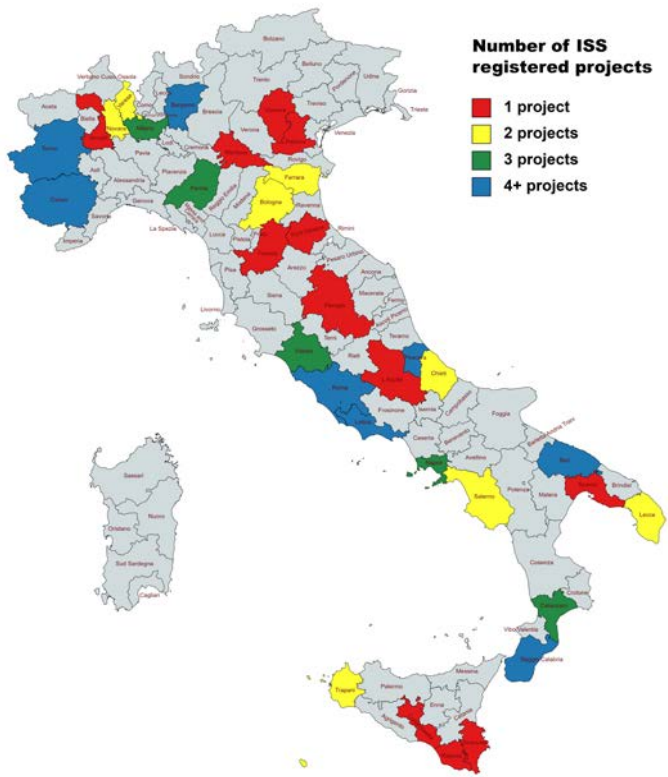

*Source*: www.mapchart.net/italy.html.

**Table 1.** *Geographic Distribution of All the Projects Registered in the ISS Contest*

| Region | Accepted projects | Proposing schools | Provinces |
|---|---|---|---|
| PIEDMONT | **43** projects | **7** schools | **4** provinces |
| LAZIO | **13** projects | **8** schools | **3** provinces |
| LOMBARDY | **10** projects | **6** schools | **4** provinces |
| CALABRIA | **8** projects | **5** schools | **2** provinces |
| EMILIA-ROMAGNA | **8** projects | **4** schools | **4** provinces |
| APULIA | **7** projects | **6** schools | **3** provinces |
| ABRUZZO | **7** projects | **5** schools | **3** provinces |
| CAMPANIA | **5** projects | **5** schools | **2** provinces |
| SICILY | **5** projects | **4** schools | **4** provinces |
| VENETO | **2** projects | **2** schools | **2** provinces |
| TUSCANY | **1** project | **1** school | **1** province |
| UMBRIA | **1** project | **1** school | **1** province |

*Source*: https://bit.ly/3ChUsus.

*The Assessment of the ISS Projects in summer 2017*

The seven members of the *ISS* evaluation committee (3 from MDI, 3 from MIUR, and a scientist as President) monitored the compliance with the rules and the merit criteria, i.e., "innovativeness, creativity and executive simplicity of the project, interdisciplinary interest, inclusion in curricula, originality of the scientific and technological content, completeness of the descriptive documentation and of the artefact / prototype / application / experiment and/or procedures, repercussions and impact in real contexts, potential developments of the project" (translated from the MIUR Note No. 10475, Attachment No.1, Italian website[10]). Let us rank in order of score the best 17 projects, out of 110 accepted submissions, selected during Summer 2017 (translated from the MIUR list, Italian website[11]):

1. 4.8 points: *GEM PBR nutrients and oxygen in space*
   Design of a machine for cultivating in microgravity some cyanobacteria able to exploit waste products.
2. 4.7 points: *Space surveyor*
   Design of a small module with a remote control system replacing the ISS astronauts during their external inspections.
3. 3.8 points: *Fitness in space: portable technology solutions*
   Training protocol using biomedical devices to ensure a correct state of health and longer time in space.
4. 3.8 points (ex-aequo): *Big brother in outer space*
   Immersive virtual reality reproducing the effects on astronauts of life in space and weightlessness.
5. 3.7 points: *Space to your breakfast*
   Study on bringing the Mediterranean breakfast aboard the Space Station.

---

[10]https://bit.ly/3vy0PW4.
[11]https://bit.ly/3vst3kX.

6. <u>3.7 points (ex-aequo)</u>: *Environmental energy signature*
   Study about energy consumption and polluting emissions in comparison with the astronaut's environment.
7. <u>3.6 points</u>: *Space Oddity*
   Design of a thermal mug to be used on the ISS.
8. <u>3.6 points (ex-aequo)</u>: *AE Space Herbs*
   Prototype to carry out hydroponic and aeroponic crops in the Space Station usable also in extreme terrestrial environments.
9. <u>3.5 points</u>: *Drop Universe*
   Study of the behavior of drops of various liquids on different materials.
10. <u>3.2 points</u>: *CRAYFIS ISS*
    Use of CRAYFIS (Cosmic RAYs Found In Smartphones) to be installed on smartphones on board the ISS.
11. <u>3.1 points</u>: *Philae and Rover Exomars conquering the deep space*
    Realization of a 1:1 scale copy of Philae and Rover Exomars integrated by an app for the remote control of the two prototypes, with particular attention to the transmission of long-distance signals (time lag).
12. <u>3.1 points (ex-aequo)</u>: *Yoga in space*
    Yoga exercises for ameliorating the astronauts' psychophysical well-being.
13. <u>3.0 points</u>: *ISS app*
    Application allowing any user to get in touch with an ISS astronaut.
14. <u>2.9 points</u>: *The dice of daily life*
    Devise a game helping the community on the ISS to choose shareable values.
15. <u>2.8 points</u>: *Gravitational rings*
    Verify an innovative type of elastic that allows exercises on board the ISS maintaining muscle mass and tone.
16. <u>2.3 points</u>: *Project Scenedesmus*
    Studying the adaptations to different environmental conditions of some algal strains (in extraterrestrial conditions or within space bases). Similar technologies could be used on Earth in extreme environments.
17. <u>2.1 points</u>: *Green Thumb*
    Cultivating some elements of the Mediterranean diet in space to improve cultivation techniques on Earth and ensure a quality diet for astronauts.

*The Announcement of the ISS Winners in Fall 2017*

On November 3, 2017, the MIUR announced the victory of seventeen projects selected by the *ISS* evaluation committee[12]; whose technology readiness level (recalling the ISO 16290:2013 standard) ranged from TRL 2 (formulation of technological applications) to TRL 5 (technology validation in relevant environment). We catalogue them by distinguishing the first place (Table 2), the second place (Table 3), and the third place (Table 4). All the projects submitted for the Area No. 4 "Observe the Earth to guard it" were ruled out and no projects classified in the third place for the Area No. 6 "Robots, satellites and astronauts conquering the Universe".

---

[12]https://bit.ly/3CsSU0G.

**Table 2.** *The First-Placed Projects of the ISS Contest for Italian Schools*

| Area | Title | School | City (Region) |
|---|---|---|---|
| 1 | *Space to your breakfast* | "Salvo D'Acquisto" | Bracciano (Lazio) |
| 2 | *Fitness in space: portable technology solutions* | "Attilio Castelli" | Saronno (Lombardy) |
| 3 | *GEM PBR nutrients and oxygen in space* | "Donatelli-Pascal" | Milan (Lombardy) |
| 5 | *Big brother in outer space.* | "Peano-Pellico" | Cuneo (Piedmont) |
| 6 | *Space surveyor* | "Augusto Righi" | Taranto (Apulia) |
| 7 | *AE Space Herbs* | "Enrico Fermi" | Mantua (Lombardy) |

*Source*: https://bit.ly/3c62Pij.

**Table 3.** *The Second-Placed Projects of the ISS Contest for Italian Schools*

| Area | Title | School | City (Region) |
|---|---|---|---|
| 1 | *Space Oddity* | "Giacomo Chilesotti" | Thiene (Veneto) |
| 2 | *Yoga in space* | "Donatelli-Pascal" | Milan (Lombardy) |
| 3 | *Drop Universe* | "Corradino D'Ascanio" | Montesilvano (Abruzzo) |
| 5 | *Environmental energy signature* | "Pitagora" | Pozzuoli (Campania) |
| 6 | *Philae and Rover Exomars conquering the deep space* | "Pietro Paleocapa" | Bergamo (Lombardy) |
| 7 | *Project Scenedesmus* | "Don Lorenzo Milani" | Romano di Lombardia (Lombardy) |

*Source*: https://bit.ly/3c62Pij.

**Table 4.** *The Third-Placed Projects of the ISS Contest for Italian Schools*

| Area | Title | School | City (Region) |
|---|---|---|---|
| 1 | *The dice of daily life* | "Umberto Pomilio" | Chieti (Abruzzo) |
| 2 | *Gravitational rings* | "Patini-Liberatore" | Castel di Sangro (Abruzzo) |
| 3 | *CRAYFIS ISS* | "Leon Battista Alberti" | Abano Terme (Veneto) |
| 5 | *ISS app* | "Antonio Meucci" | Ronciglione (Lazio) |
| 7 | *Green Thumb* | "Publio Virgilio Marone" | Meta (Campania) |

*Source*: https://bit.ly/3c62Pij.

From Tables 2–4 we can visualize the regional distribution of the selected schools on the Italian territory as follows (Table 5 and Figure 7).

**Table 5.** *Geographic Distribution of the Italian Schools Winning the ISS Contest*

| Region | First place | Second place | Third place |
|---|---|---|---|
| LOMBARDY | **3** projects | **3** projects | ------------------------- |
| LAZIO | **1** project | ------------------------- | **1** project |
| APULIA | **1** project | ------------------------- | ------------------------- |
| PIEDMONT | **1** project | ------------------------- | ------------------------- |
| ABRUZZO | ------------------------- | **1** project | **2** projects |
| CAMPANIA | ------------------------- | **1** project | **1** project |
| VENETO | ------------------------- | **1** project | **1** project |

**Figure 7.** *The 2017 Italian Regional Map of the ISS Winning Projects*

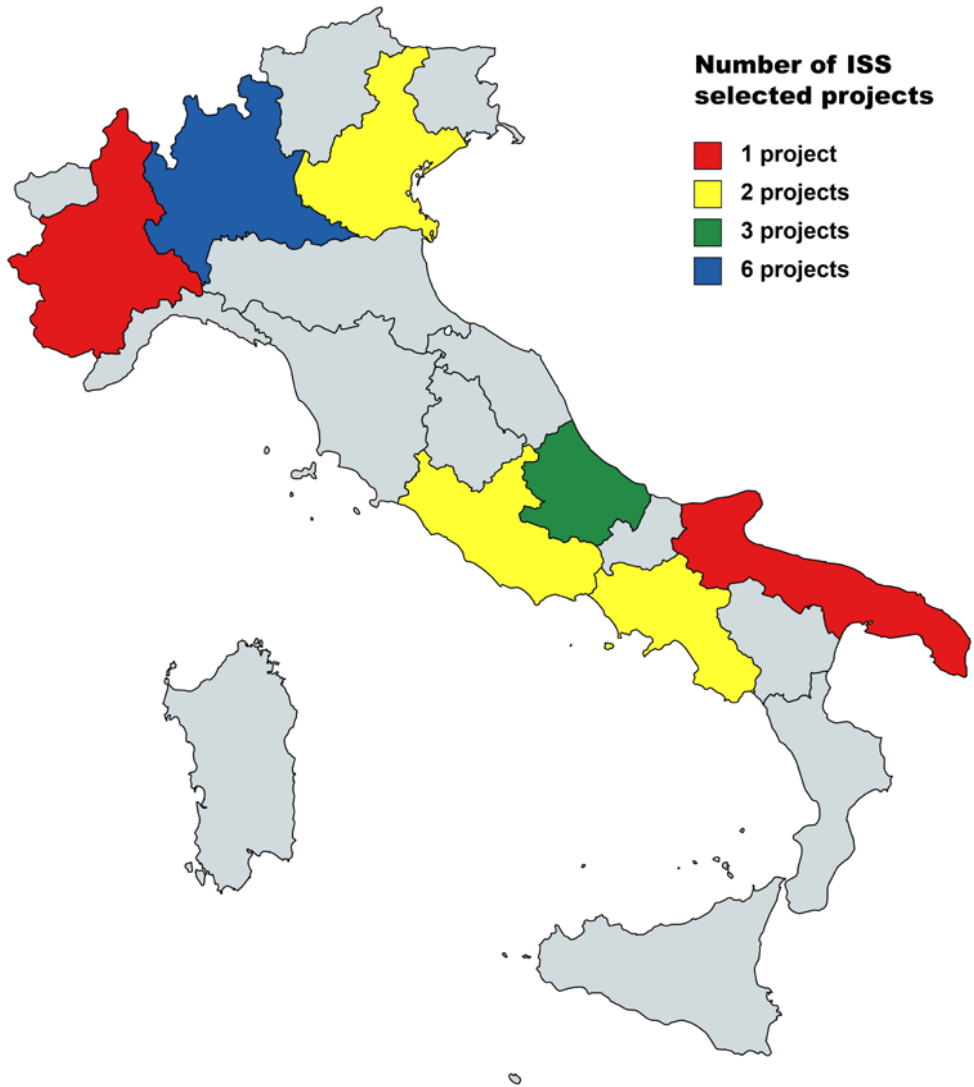

*Source*: www.mapchart.net/europe-detailed.html.

The peaks of performance in the most productive parts of Italy, such as the whole Lombardy region and the industrial zone of Taranto, confirm that both STEM career knowledge and interests are influenced by society at large (Blotnicky et al., 2018) and benefit from the partnerships with industry (Nistor et al., 2018) and multi-stakeholders (Jiménez-Iglesias et al., 2016). MIUR embraced the synergy of all these factors when introducing the: "technical tutorship a further element of originality of the initiative consists in the provision of a tutorship mechanism for the participating schools by third parties with specific skills in the fields contemplated by the call for applications, such as research centres, industries, universities, etc. On the one hand, this collaboration will offer enrichment opportunities for schools according to the *learning by doing* and *skills-based teaching* methodology, and on the other hand it will create local synergies, combining *production* and *training*. Thus, starting from November 2016 and up to March 2017, the preliminary projects will be developed and implemented with the tutoring of one or more partners (company, authority, foundation, institution, industrial district, etc.) identified among the suitable subjects of the area and indicated in the preliminary project. The schools can activate contacts with the subjects indicated above autonomously or during in-depth workshops. The responsibility for identifying, assessing, and accepting the third-party partnership is up to the participating school." (Translated from the MIUR Note No. 10475, Italian website[13]).

*The ISS Awarding Ceremony in winter 2018*

On March 14, 2018, the winning schools were awarded at the auditorium of the Maxxi Museum in Rome. The event was attended by the Deputy Chief of Staff of the Ministry of Defense (Brigadier General Maurizio Cantiello), the Head of the 5th Department of the Air Force General Staff (Division General Giorgio Baldacci), and representatives of the Italian Space Agency and Thales Alenia Space. Cosmonaut Lieutenant Colonel Walter Villadei presented the works on behalf of the Selection Committee (ASI TV[14]). The celebration, introduced by the scientific communicator of Rai Cultura, Davide Coero Borga, was also attended by the Italian Minister of Education, University and Research, Valeria Fedeli, who declared: "Dear students, dear all, I am particularly happy to be here today for the awards ceremony of the national competition *School: space for your future. The ISS: Innovatio, Scientia, Sapientia*, which the Ministry of Education, University and Research promoted together with the Ministry of Defense and in collaboration with the Italian Space Agency. I am glad not only because we reward the ideas, talent, and intelligence of the new generations, but also because this competition was an extraordinary occasion of knowledge, of *school beyond school*, which gave the opportunity to female and male students to deepen the internal mechanisms of research, work, and future sectors. And they did so thanks to the support of experts, who have put their skills and professionalism at the service of young people. I want to thank all those who have been partners of this initiative: universities, research institutes, private companies, associations. Synergy in education is essential if we want to offer the girls and boys who attend our institutes opportunities for free and healthy growth. Why did we want a contest like this? A competition prompting female and male students to make proposals that could even be considered by ASI among the activities that will take place on board the International Space Station (ISS). Through this competition we wanted to contribute to the cultural growth and training of the younger generations, strengthening their knowledge and skills in the scientific and engineering disciplines; increase the interest of the school community for space and aerospace research; generate synergies and initiatives capable of increasing Italian participation, in the future, in a highly technological and rapidly expanding sector. The young participants – to whom I address my compliments – were able to challenge themselves by confronting and developing scientific and technological experimentation projects within seven thematic areas: from daily life to Space; train like an astronaut; test the Sciences in Space; observe the Earth to guard it; stay connected with an astronaut; robots, satellites and astronauts conquering the Universe; cultivate in space to cultivate better on Earth. Original and very interesting proposals have arrived: from the proposal of exercises to improve the psycho-physical health of astronauts through yoga to the study of how to bring the Mediterranean breakfast into space on board the Space Station. Or, ulteriorly, from the application allowing any user to get in touch with an astronaut on the International Space Station to the project that plans to cultivate some of the elements making up the Mediterranean diet in space to improve cultivation techniques on Earth and ensure astronauts a quality food diet. Today we reward these ideas, these works, the result of research, study, and in-depth analysis. But, at the same time, we break down a barrier, we put in communication worlds that may appear distant but are not (and must not be). We tell the new generations that any experience, even the most imaginative one, requires commitment, study, work and in some cases even sacrifice. That any profession, especially one of such great responsibility, requires solid skills and knowledge. Which are the inextinguishable thirst for knowledge and the determination to constantly test ourselves to take us into tomorrow, in the future, as in space. Space is marking, and increasingly will, the rhythm of our social and economic life. We have to look at this as a *system of systems*, based on the integration of different structures, technologies, and services, both *terrestrial* – traditionally understood – and typical of space programs. Our aim is to transform the European space sector into one of the driving forces for growth. We must see it as an engine and integrator of technological innovation processes in society as a whole. We must remember the great contribution that Space can make in the formation of the new generations by stimulating them to undertake a scientific career and letting them grow a strong sense of European identity. Only in this way our young people will be the protagonists of the new economic, political, and anthropological balances that are taking shape today. Paolo Nespoli, an astronaut with invaluable experience, recently returned from the VITA mission, has said: *"Being an astronaut on a rocket that is about to leave is like sitting on a mini atomic bomb that is about to explode in a controlled way and then if all goes well it throws you into space"*. Dealing with a space mission is anything but simple. It is a task requiring study. But it is also, at the same time, an important chance for progress and development. From today to the next 10 years the panorama will change, showing us new scientific and technological challenges, with further repercussions for Europe. It is an opportunity we cannot miss: we can grow, and we can do it together, thanks to the contribution of all those involved in this process. The exploration of space is an experience serving as an example and a reference, both practical and metaphorical, to express that need to discover, that curiosity to investigate that drive to go into the unknown to make it familiar. Isn't it, perhaps, the basis of knowledge and wisdom in general? The desire to overcome limits and boundaries, to cross new terrains different from their usual ones and to assume different perspectives, if not diametrically opposed, to the traditional ones? What brought us into Space is the same drive to learn that makes knowledge grow, that motivates us to study, that moves the world towards new horizons. A push that is not only stemming from an irrational and uncontrolled impulse but is rather the result of a study and an in-depth analysis, that are essential conditions for approaching unknown realities. Dear girls and boys, take this cognitive experience with you, put it to good use. Keep it as track of each of your studies, professional and life paths. There are no boundaries that you cannot cross if you commit yourself, if you have the strength to believe in your dreams, if you have the courage to put yourself to the test. You have proved it to yourself thanks to this contest. We are with you." (Translated from the Letter of the Minister Fedeli to the *ISS* winners, Italian website[15]).

---

[13] https://bit.ly/3Pv0haY.
[14] https://bit.ly/3KfOMDo.

*The Aftermath*

Up to date, none of the seventeen *ISS* winning projects appears on the "Complete list of the Italian microgravity experiments" (ASI website[16]). It seems implausible that secondary schools' proposals could be pondered, even just as basic insight, for being "subsequently developed and engineered, according to the necessary requirements, for experimentation in space flight" (translated from the MIUR Note No. 10475). Although missing its highest remuneration, the *ISS* stepped a milestone in the history of Italian Education; the spontaneous response from 12 regions (out of 20) and 33 provinces (out of 107) induced the Italian Ministry of Education to continue promoting Astronomy in STEM classrooms[17] by signing collaborations with:

- the Italian Astronomical Society (SAIt), aimed at disseminating the scientific culture and the sky sciences at school (April 9, 2021[18]);
- the European Space Agency (ESA), oriented to space-based innovation and digitalization for the school of tomorrow (June 21, 2021[19]).

In addition, the mentoring role was institutionalized on December 22, 2022[20], by assigning a *Tutor* to each class group; this novel student assistance plan[21] corroborates the long tradition of *inclusivity* of the Italian school (Ianes et al. 2020).

*Educational Impact*

The "Space for Your Future" was universally perceived as a rare incentive to contrast the low attractiveness of STEM studies (Nistor et al., 2018) through the STEM best practices (Sanders, 2012). In fact, all the presented projects were accomplished via PBL & PrBL (Barron et al., 1998) and boosted the IBSE (Muciaccia et al., 2019), the ESD (Gras-Velázquez & Fronza, 2020) and, in many cases, the EDP (Hafiz & Ayop, 2019) and the ICT (Gras-Velázquez, 2016, 2017). Eventually, the "Innovatio, Scientia, Sapientia" contributed to reduce the gender gap in Science for secondary students (Makarova et al., 2019), to activate the emotional intelligence (Parker et al., 2004), and to reinforce the use of Science Laboratories (Hofstein & Mamlok-Naaman, 2007), eminently the Chemistry Lab (Hofstein, 2004). The entire *ISS* experience and the two-fold contribution from the "G.B. Grassi" of Latina, the scientific high school hosting the Planetarium "Livio Gratton" (Bonacci, 2011a), raised a genuine interest in the Section "Didactics of Physics" of the 104th Congress of the Italian Physical Society (September 17–21, 2018) at the University of Calabria – UniCal (Bonacci, 2018a) and in the Theme "BE a citizEn Scientist (BEES)" of the 13th European Researchers' Night (September 28, 2018) by Frascati Scienza (Bonacci, 2018b). The latter was the keynote lecture of "The Museum communicates and demonstrates Science", a citizen science meeting held at the "Museo della Terra Pontina"[22] (translatable as "Museum of the Pontine Land"), in the city of Latina (Miltiadis, 2020), and based on classroom works (Nistor et al., 2019).

---

[15] https://bit.ly/3Pzw1Mg.
[16] https://bit.ly/3R59LL7.
[17] https://bit.ly/3Hr7FSF.
[18] https://bit.ly/3HlSryo.
[19] https://bit.ly/3wTVkBH.
[20] https://bit.ly/3qKn9x0.
[21] https://bit.ly/3qILfZ3.

**The ISS Educational Project by the "G.B. Grassi" High School**

We focus on the educational project "Space for your Future. The ISS: Innovatio, Scientia Sapientia" by which the Scientific High School "Giovanni Battista Grassi" of Latina participated in the homonymous call for applications proposed by the Ministries of Education and Defense in the years 2016–2017. The Province of Latina has a distinct agricultural propensity, with the largest number of farm workers in the Lazio region (slightly higher than Rome)[23]. Latina is also the second province (next to Milan) for number of employees in pharma companies, and the first for pharmaceutical exports (ahead of Milan)[24]. The initiative was sponsored in Latina by the cosmonaut Walter Villadei who convinced the author, as Director of the Planetarium "Livio Gratton" (Bonacci, 2013), to participate with an *ad hoc* project based on *constructivism*. Starting from October 2016, ten last-year pupils were divided into two groups of five (Team A & B) who developed the following experiences to be tested in space:

A. "Chemical transformations with precipitate formation",
B. "Redox reaction of colored reagents".

In both cases, the goal was the same: spotting the differences between the terrestrial and the space (micro-g) environment about the evolution of a low technology readiness level (TRL 4) chemical reaction. Owing to the protracted duration of their reactions, Teams A & B excluded ground-based drop facilities, like the *Einstein-Elevator*[25], and any other experimental setup for domestic microgravity alternative to the International Space Station.

---

[22]https://bit.ly/3RKTtIX.
[23]https://bit.ly/3Hwnskw.
[24]https://bit.ly/3jbez6v.

*Project's Steps*

The school activity developed through the following phases:
1. September 2016: the School Council approved the *ISS* project presented by the author, who nominated Francesco Giuliano, Provincial Head of the International Year of Astronomy (IYA09) and of Chemistry (IYA11), as technical tutor for his curriculum of skilled chemist (Giuliano, 2008), appreciated teacher (Giuliano, 2011), and fine pedagogist (Giuliano, 2015).
2. October 2016: the author selected 10 motivated pupils who familiarized themselves with the updated regulation of the competition, studying the ISS as the home of humanity in space, and the significant contribution Italy gave to its construction (ASI website[26]).
3. October 2016 – November 2016: the students chose the III thematic area "Test the Sciences in Space", absorbing the concept of microgravity[27] and contriving two germinal projects (here identified as *ISS–A* and *ISS–B*).
4. November 7, 2016: the Principal Giovanna Bellardini sent the preliminary projects *ISS–A* and *ISS–B* to MIUR by registered mail.
5. November 2016 – January 2017: the pilot projects were refined via *learning by doing* (Dewey, 1916) in the Chemistry Lab (TRL 4).
6. January 2017 – March 2017: the experimental proposals were engineered through PBL & PrBL (Zhou et al., 2011) and EDP (Lin et al., 2021).
7. March 21, 2017: the definitive projects *ISS–A* and *ISS–B* were submitted to the Directorate General for the Regulations and Evaluation of the National Education System of MIUR (Bonacci et al., 2017a, 2017b).

They obtained a "moral victory", for being amid the few teams able to send their *final* results in time. In fact, the deadline for the submission was unexpectedly prolonged from March 31 to May 31, 2017.

*Project's Targets*

STEM-focused activities require at least two different subjects (Kelley & Knowles, 2016), in our case: Chemistry and Astronomy. They are both curricular in the Italian schools, so there was no need for any special training. The *ISS* project studied simple chemical transformations, in aqueous system, validated in lab (technology level TRL 4); precisely, the calcium carbonate precipitation and the oxidation-reduction between copper sulfate and iron. The pupils used common substances, easily available and not harmful, noting down their physical properties and preparing the relative solutions in distilled water, whose concentrations were defined in the execution phase. Appendixes A and B provide additional details about the experimental tests.

---

[25] https://bit.ly/3p0Xqzl.
[26] https://bit.ly/3R8SmS4.
[27] https://bit.ly/3xSkukD.

*Project's Strategies*

The project manager and the technical tutor adopted the constructivist methodology (Piaget, 1954) letting the high schoolers develop organization and problem-solving abilities. For each test, the students annotated their observations in forms drawn up under the teachers' guidance. These forms were also prepared for the astronauts who, in the event of victory of one or both the proposed projects, would have compiled them with the experimental results obtained in the *micro-g* environment.

*Project's Achievements*

The *ISS* educational project contributed to reach the benchmarks of STEM readiness (Mattern et al., 2015), improving the STEM competences (Boon Ng, 2019) beyond the classic 4C skills (Triana et al., 2020). Both the Teams A and B practiced the PBL & PrBL (Smith et al., 2022) and the EDP (English & King, 2015) within a *learning by doing* pattern based on feeding the students' curiosity (Kowalski & Kowalski, 2015). Guided by two male supervisors, the brilliant performance of Team B (with four female members out of five) showed that the gender gap in STEM fields is not necessarily due to mismatching mentors, as conversely claimed by Kricorian et al. (2020) and by Bilgin et al. (2022a). Nine of the ten *ISS* high schoolers enrolled in a STEM faculty in Fall 2017; it was an impressive percentage compared to the average 25% of STEM graduated in OECD countries until then (OECD, 2017), albeit not representing a statistical sample. Besides, all the six *ISS* female pupils succeeded in a STEM major; a stunning performance compared, e.g., to the 26% of UK young women graduated in core STEM subjects in 2019[28].

**Didactical Summary**

The Italian schools' competition "Innovatio, Scientia, Sapientia" contained all the driving factors for implementing a STEM pedagogy of that *high-quality* pursued by the European Education Area (EEA)[29], and confirmed *Astronomy* as a gateway science[30] that should be inserted in any school curriculum (Percy, 2005). By virtue of fully sustainable projects (Carroll et al., 2019) a broad spectrum of STEM skills[31] was covered: problem solving, creativity, critical analysis, teamwork, independent thinking, initiative, communication, and digital literacy. The *ISS* optimized the STEM good practices (Kasza & Slater, 2017), by means of:

---

[28]https://bit.ly/3R8lSbC.
[29]https://bit.ly/3L3Z7Vj.
[30]https://bit.ly/3GYXTqF.
[31]https://bit.ly/3ZgWW5D.

the PBL & PrBL (Wilhelm et al., 2023), the IBSE (EUN, 2019), the ESD (Bilgin et al., 2022b), the EDP (Winarno et al., 2020), and the ICT (Lukychova et al., 2021). The feedback of several secondary schools, such as the "G.B. Grassi" of Latina (big city in Lazio), revealed the importance of using the Science Laboratories (Hofstein & Kind, 2012), and of nourishing the students' curiosity (Fetto, 2015), by activating their emotional intelligence (Petrides et al., 2004). In this perspective, the memorandums of understanding between MIUR and, respectively, SAIt and ESA, are comforting leaps forward.

**Conclusions**

This paper is aimed at bolstering the current European Schoolnet's debate on STEM education[32] and disseminating Educational Science, Innovation and Research in the wake of the erstwhile project DESIRE[33]. We have delineated the "Space for Your Future. The ISS: Innovatio, Scientia, Sapientia", a STEM enrichment activity inheriting the legacy of the SUCCESS contest by ESA[34]. The *ISS* was a space competition promoted by MIUR & MDI in the years 2016–2017 to select, after strict scrutiny, experiments suitable for the International Space Station. It was massively participated, with 110 proposals presented by 54 Italian secondary schools. From the complete list of the Italian microgravity experiments by ASI we infer that neither one of the 17 winning projects has been considered yet for testing on board the International Space Station. Nevertheless, the "Innovatio, Scientia, Sapientia" was a formidable stimulus for STEM tasks accomplished by teen students throughout Italy (TRL 2–5). Indeed, the collective effort to supply new ideas for space exploration involved at least another subject cognate to Astronomy, such as Mathematics, Physics, Chemistry, Biology etc. The correlation between the industrialization of a land and the positive response by territorial schools was confirmed in Latina, a manufacturing Italian province where the Scientific High School "Giovanni Battista Grassi" joined the *ISS* challenge with two teams. They participated through a *learning by doing* model (supremely via PBL & PrBL) leading to portable solutions in the field of Chemistry. Namely, they proposed to test in space a redox reaction of colored reagents and a chemical transformation with precipitate formation, each in an appropriately designed reaction vessel. The high schoolers accepted the "Space for Your Future" with enthusiasm also because, in fall 2016, the "G.B. Grassi" had already a consolidated experience of STEM projects and citizen science around its Planetarium "Livio Gratton". Other success factors were the availability of a well-equipped chemistry laboratory at school and the technical tutorship of Francesco Giuliano, an excellent constructivist teacher of Chemistry. Being among the few schools which submitted their results within the first deadline of March 31, 2017, the "G.B. Grassi" obtained a *moral victory* and its *ISS* projects were popularized in major conferences. Emerging as pivotal of superb STEM pedagogy, *Astronomy* should be included in every school's curriculum as single subject or interdisciplinary course and a permanent partnering with national or international space agencies should become a priority of any institution meeting with the EEA's standards for a quality education. In our opinion, exemplary Stem & Space programs are *NAC*[35], *EEE*[36], *S4G*[37], *MoCRiS*[38] for school pupils, and *Micro-g NExT*[39], *DEPLOY!*[40], *PETRI*[41] for undergraduates.

---

[32]https://bit.ly/3RfTqng.
[33]https://bit.ly/3hXNwes.
[34]https://bit.ly/3KfsPGc.

**Remarks**

In 2020, the Italian Ministry of Education, University and Research (MIUR) split in two:

1) the Ministry of Education (MI)
2) the Ministry of University and Research (MUR)

The Italian Ministry of Education (MI), responsible for primary and secondary schools, further changed into the Ministry of Education and Merit (MIM) on November 4, 2022[42].

**Acknowledgments**

We are grateful to Francesco Giuliano for his voluntary participation as precious tutor in the *ISS* educational project by the Scientific High School "G.B. Grassi" of Latina. We also thank the anonymous reviewers for their helpful comments and valuable suggestions.

---

[35] https://bit.ly/3qsU3l2.
[36] https://bit.ly/43PprbS.
[37] https://bit.ly/3Ct5LPf.
[38] https://bit.ly/3qPNBVD.
[39] https://go.nasa.gov/42Mn34J.
[40] https://bit.ly/46eaO41.
[41] https://bit.ly/3JbfVrF.
[42] https://bit.ly/3XtoyTP.

**Appendix A – Precipitation Reaction**

The Project *ISS–A* (Table 6) verifies the precipitation of calcium carbonate in *microgravity*, comparing it with the same reaction performed on Earth: $Ca(OH)_2$ (aq) + $CO_2$ (g) = $CaCO_3$ (s) + $H_2O$ (l).

**Table 6.** *The ISS Project A of the Scientific High School "G.B Grassi"*

| Title | Chemical transformations with formation of precipitate |
|---|---|
| **Reaction** | $Ca(OH)_2$ (aq) + $CO_2$ (g) = $CaCO_3$ (s) + $H_2O$ (l) |
| **Materials** | Mass $Ca(OH)_2$: m = 0.4 g, Volume $H_2O$: V = 300 mL (t=20° C) |
| **Tools** | 400 mL beaker, balance (sensitivity 0.01 g), thermometer, straw |
| **Area** | No. 3 "Test the Sciences in Space" |
| **Project Manager** | Enzo BONACCI, Teacher of Mathematics and Physics, Director of the Planetarium "Livio Gratton" in Latina |
| **Students** | Gabriele CALDARINI, Francesca Romana CALVI, Riccardo CATANIA, Simone CIOTTI, Alessandra DEL MONTE |
| **Project external Tutor** | Francesco GIULIANO, retired Teacher of Chemistry, former Lecturer of Chemistry Didactics at SSIS (School of Specialization in Secondary Teaching), Province Head of IYA09 & IYC11 in Latina |

*Source*: https://bit.ly/3B0zMq1.

The Team A mixed 0.4 grams of calcium hydroxide in 300 milliliters of distilled water until it reached transparency in the beaker. After blowing into the straw (i.e., adding carbon dioxide), the solution became *opaque* in 2 minutes and 30 seconds. The solute *precipitated* completely in circa 8 hours (Fig. 8). Since the terminal velocity of a sphere falling in a fluid is directly proportional to the acceleration due to *gravity*, according to Stokes' law, the Team A wished the ISS crew to reply to the following question:

1. Will $CaCO_3$ precipitate in space or will the solid remain suspended?

**Figure 8.** *From the Opaque Solution to the Full Precipitation of $CaCO_3$*

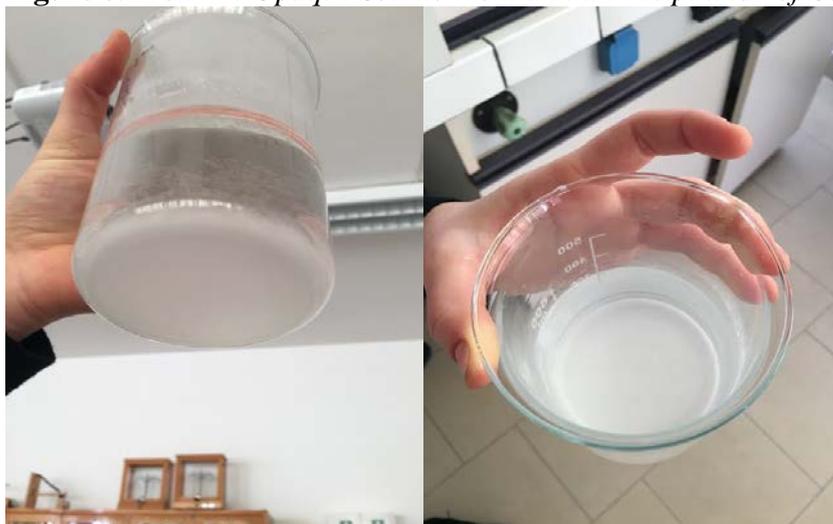

*Source*: https://bit.ly/3B0zMq1.

The Team A excogitated a transparent, inert acrylic *reaction vessel* fitting the conditions aboard the International Space Station (Fig. 9), with a *red* flow valve and a *green* external/internal pressure balancing valve.

**Figure 9.** *Front View of Team A's Vessel 3D Model*

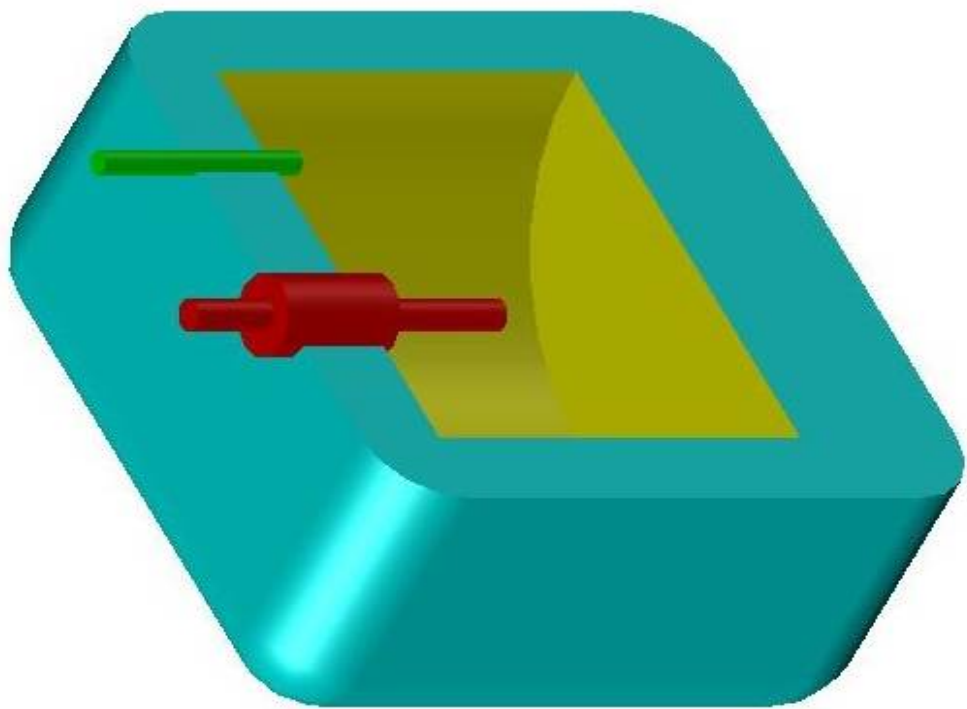

*Source*: https://bit.ly/3B0zMq1.

The ISS astronauts should blow on the mouthpiece of the *red* flow valve, measure the interval of time for an opaque solution and, if it happens, for the precipitation of $CaCO_3$; next they should fill in the two missing entries in the following Table 7.

**Table 7.** *The Astronauts' Form for the Project ISS–A*

| Test's data and results | Experiment on Earth | Experiment on the ISS |
|---|---|---|
| Volume of $H_2O$ | 300 mL (at t=20° C) | 300 mL (at t=20° C) |
| Mass of $Ca(OH)_2$ | 0.4 g | 0.4 g |
| Solubility of $Ca(OH)_2$ | 1.7 g/L (at 20°C) | 1.7 g/L (at 20°C) |
| Time of opacification | 2 minutes and 30 seconds | |
| Time of precipitation | 8 hours | |

*Source*: https://bit.ly/3B0zMq1.

The main merits of the *ISS–A* and *ISS–B* are a clear feasibility and a striking inventiveness, being projects never tried in ESA's educational activities[43] and programmes[44].

---

[43]https://bit.ly/43xlskF.
[44]https://bit.ly/43wFP1m.

**Appendix B – Redox Reaction**

The Project *ISS–B* (Table 8) verifies the oxidation-reduction between copper sulfate and iron in *microgravity*, comparing it with the same reaction performed on Earth: Fe (s) + $Cu^{2+}$ (aq) → $Fe^{2+}$ (aq) + Cu (s).

**Table 8.** *The ISS Project B of the Scientific High School "G.B Grassi"*

| Title | Redox reaction of colored reagents |
|---|---|
| Reaction | Fe (s) + $Cu^{2+}$ (aq) → $Fe^{2+}$ (aq) + Cu (s) |
| Materials | 55 mL of $H_2O$, 15.8 g of $CuSO_4$, 3.53 g of iron filings (Fe) |
| Tools | 150 mL beaker, balance (sensitivity 0.01 g), thermometer |
| Area | No. 3 "Test the Sciences in Space" |
| Project Manager | Enzo BONACCI, Teacher of Mathematics and Physics, Director of the Planetarium "Livio Gratton" in Latina |
| Students | Giulia CHILLEMI, Federica MADDALONI, Alice PACINI, Gianluca SBANDI, Chiara TRUINI |
| Project external Tutor | Francesco GIULIANO, retired Teacher of Chemistry, former Lecturer of Chemistry Didactics at SSIS (School of Specialization in Secondary Teaching), Province Head of IYA09 & IYC11 in Latina |

*Source*: https://bit.ly/3B0zMq1.

The Team B mixed, in the beaker, 15.8 grams of copper sulphate ($CuSO_4$) in 55 milliliters of distilled water until it reached a *blue* color at 18°C. They add 3.53 grams of iron filings (Fe) and the solution became *dark grey*, turning *dark red* as copper precipitated (Fig. 10). After stirring, the *green* ferrous sulphate ($FeSO_4$) formed in circa 4 minutes and the temperature reached 33°C. The Team B wished the ISS crew to answer to the following question:

1. Will the redox in space evolve in the same way as far as reaction time, product color, and temperature are concerned?

**Figure 10.** *Reaction Between $CuSO_4$ and Fe Until the Formation of $FeSO_4$*

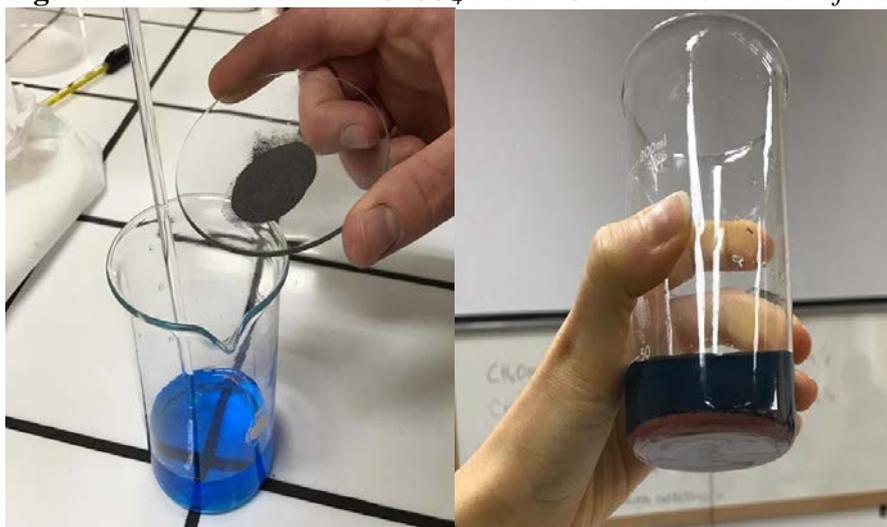

*Source*: https://bit.ly/3B0zMq1.

The Team B devised a transparent, inert acrylic *reaction vessel* fitting the conditions aboard the International Space Station (Fig. 11), with two lateral buttons for introducing the copper sulphate (first push) and the iron filings (second push), and a central winged mixer.

**Figure 11.** *Top View of Team B's Vessel 3D Model*

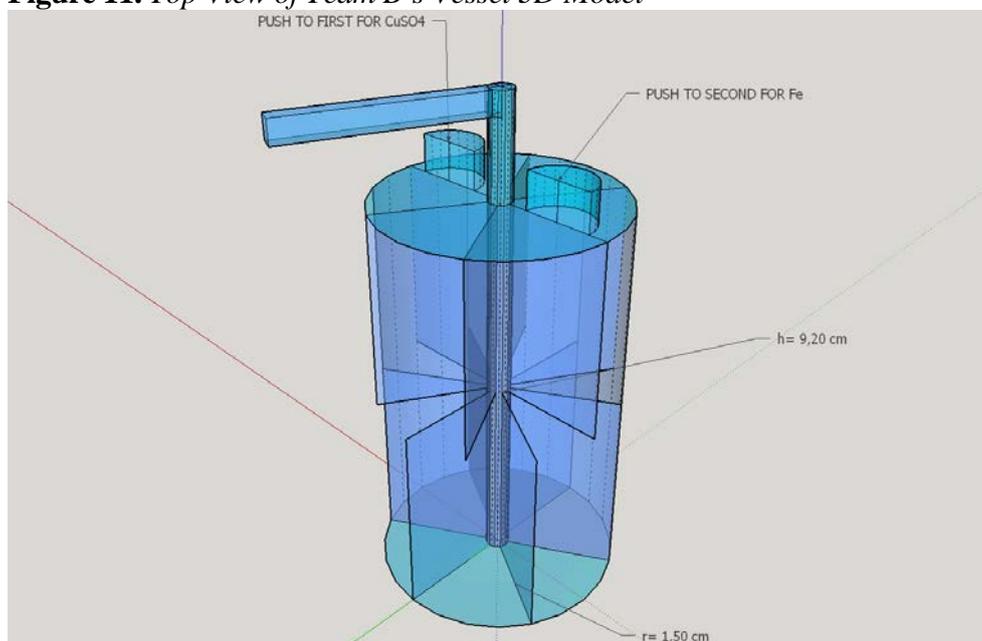

*Source*: https://bit.ly/3B0zMq1.

The ISS astronauts are expected to push the *$CuSO_4$* button and fill the first two lines of the last column (Tab. 9). Then they should push the *Fe* button and fill the third and fourth lines of the form. They have to rotate the mixer, waiting for the $FeSO_4$ formation, and complete the form. If measuring the temperature was too difficult on board the ISS, the empty cells (second and seventh lines) should be left blank.

**Table 9.** *The Astronauts' Form for the Project ISS–B*

| Test's data and steps | Experiment on Earth | Experiment on the ISS |
|---|---|---|
| Color after adding $CuSO_4$ to water | Blue | |
| Initial temperature | 18 °C | |
| Color soon after adding iron filings | Dark grey | |
| Color during copper's precipitation | Dark Red | |
| Color after mixing of reagents | Green | |
| Period of $FeSO_4$ formation | 4 minutes | |
| Final temperature | 33 °C | |

*Source*: https://bit.ly/3B0zMq1.